\documentclass[pdftex,prl,superscriptaddress,showpacs,twocolumn]{revtex4}

\pdfoutput=1

\usepackage[pdftex]{graphicx,color}
\usepackage[pdftex]{hyperref}

\usepackage{amsmath,amssymb} 

 \newcommand{\ket}[1]{|{#1}\rangle}

\newcommand\ketbra[2]{| #1 \rangle\!\langle #2 |}

\newcommand{\Ham}{\mathcal{H}}

\begin{document}
    
\title{Extending Quantum Coherence in Diamond}

\author{C. A. Ryan }
\affiliation{Department of Nuclear Science and Engineering, Massachusetts Institute of Technology, Cambridge, Massachusetts 02139, USA }

\author{J. S. Hodges}
\affiliation{Department of Electrical Engineering, Columbia University, New York, NY 10027, USA. }

\author{D. G. Cory }
\affiliation{Department of Nuclear Science and Engineering, Massachusetts Institute of Technology, Cambridge, Massachusetts 02139, USA }
\affiliation{Institute for Quantum Computing and Department of Chemistry, University of Waterloo, ON, N2L 3G1, Canada.}
\affiliation{Perimeter Institute for Theoretical Physics, Waterloo, ON, N2J 2W9, Canada}

\begin{abstract}
We experimentally demonstrate over two orders of magnitude increase in the coherence time of nitrogen vacancy centres in diamond by implementing decoupling techniques.   We show that equal pulse spacing decoupling performs just as well as non-periodic Uhrig decoupling and has the additional benefit that it allows us to take advantage of ``revivals" in the echo (due to the coherent nature of the bath) to explore the longest coherence times.    At short times, we can extend the coherence of particular quantum states out from $T_2^*=2.7\mu s$ out to an effective $T_2 > 340\mu$s.   For preserving arbitrary states we show the experimental importance of using pulse sequences, that through judicious choice of the phase of the pulses, compensate the imperfections of individual pulses for all input states.    At longer times we use these compensated sequences to enhance the echo revivals and show a coherence time of over 1.6ms in ultra-pure natural abundance $^{13}C$ diamond. 
\end{abstract}  


\maketitle

The nitrogen-vacancy (NV$^{-}$) centre in diamond is a model quantum system with coherence times of milliseconds, nanosecond gate times, and an optical handle to allow initialization and readout of single centres \cite{Wrachtrup:2006p234}. The extraordinary coherence time is essential to proposals for using NV centres in quantum information processing (QIP)  \cite{Cappellaro:2009p1554} or magnetometry  \cite{Degen:2008p1448,Taylor:2008p1323}.  However, the long coherence time of the NV defect is not immediately exploitable;  it is necessary to decouple the electron spin from unwanted interactions with its spin-based environment, that would otherwise lead to a few $\mu$s decay time.    Decoupling techniques -- applying periodic control pulses to suppress the interactions -- provide a well understood solution leveraging decades of use in magnetic resonance.   

The dominant dephasing mechanism of the NV centre in high-purity diamond is the surrounding spin bath of  $^{13}C$ nuclei in the crystal lattice \cite{Maze:2008p1637}.  The NV is an effective spin 1 but a static magnetic field lifts the $m_s=\pm1$  degeneracy and we can concentrate on the effective two-level qubit system of the $m_s=0/1$ states.     With the magnetic field along the NV symmetry axis, a secular approximation for the effective two-level NV Hamiltonian is given by:
 \begin{equation}
 \label{NVHam}
 \mathcal{H}_{NV} =  \omega_S S_z + \sum_j\omega_jI_z^j + S_z\sum_jA_j\cdot \vec{I^j} + \Ham_{dip.},
 \end{equation}
where $S/I^j$ are electron/nuclear spin operators; $A_j$ the hyperfine coupling to the $j$'th nuclei; and $\Ham_{dip.}$ the dipolar coupling between nuclei.  An electron spin superposition state is dephased by time-variations in the $S_z$ operator from both fluctuations in the external field and entanglement with the nuclear bath.   At room temperature there is an additional incoherent decay when we average over all nuclear configurations of the initial maximally mixed state during the $\sim10^6$ experiment repetitions.    

\begin{figure}[htbp]
\begin{tabular}{c}
\includegraphics[scale=0.55]{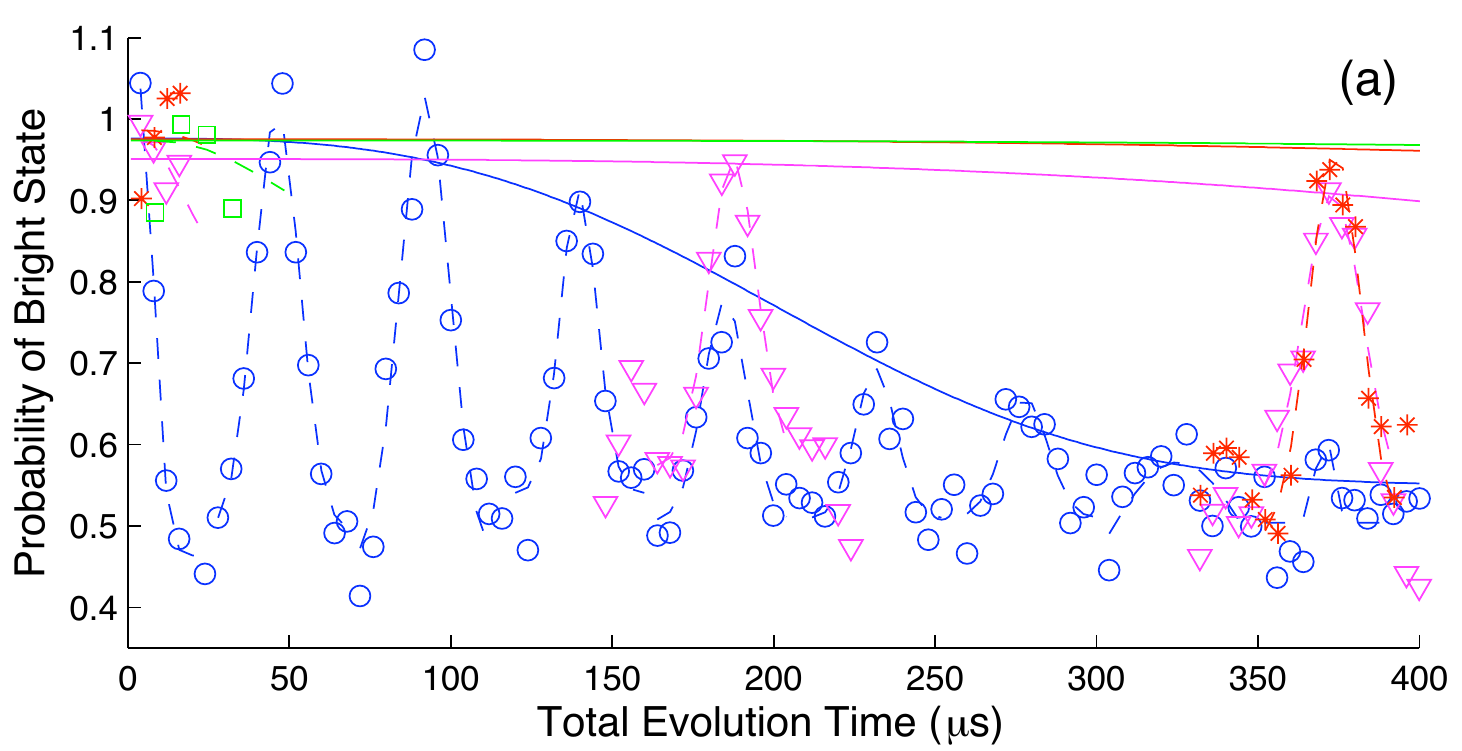} \\
\includegraphics[scale=0.55]{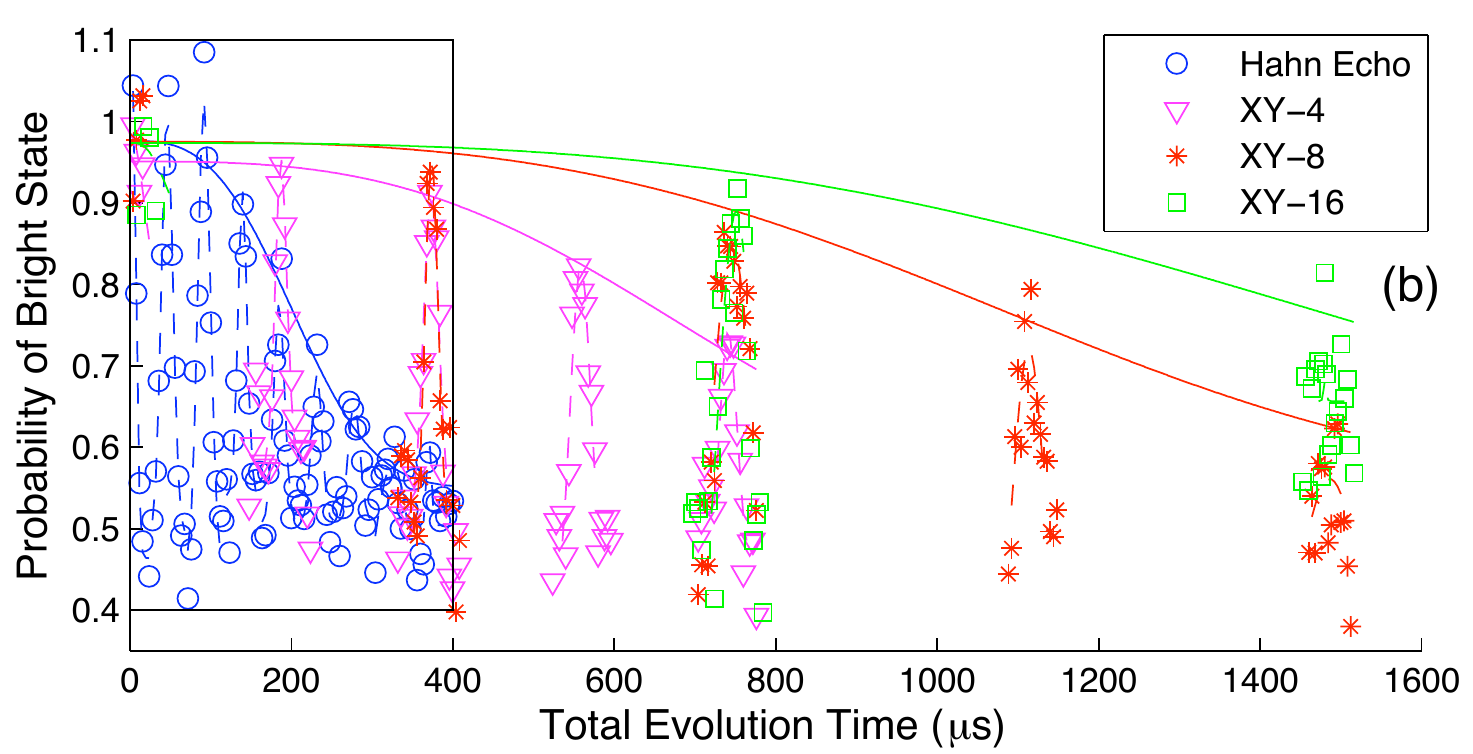} 
\end{tabular}
\caption{\label{RevivalsExpData} The effect the robust XY multiple pulse sequences on the revivals at short times (a) and longer times (b).  At short times the periodic revivals of the single spin echo are observed to decay with at $220\, \mu s\, T_2$.  With multiple pulses, the revival frequency is reduced proportionally but the $T_2$ is greatly extended to over 1.6 ms. The envelopes are fit to $s(t) = 0.5 + A\exp(-(\frac{t}{T_2})^3)$.   Since the revivals only occur at certain multiples of the bare $^{13}C$ frequency, we observe only at these times to reduce experimental time.  }
\end{figure}

Because these fluctuations are relatively slow, a spin-echo (a $\pi$-pulse with equal delays, $\tau$, before and after) can reverse some of the evolution.  However, a key part of the dynamics is that the quantization axis of a $^{13}C$ spin depends on the state of the electron, due to the anisotropic form of the hyperfine interaction \cite{hodges:010303}; this gives collapses and revivals in the electron coherence  \cite{L.Childress10132006}.  Because there are no $S_x$ and $S_y$ terms  in the secular Hamiltonian, for a free-evolution time $\tau$, the unitary propagator for the electron-bath system is $U_{e,b} = \ketbra{0}{0}_e\otimes U_{b,0} + \ketbra{1}{1}_eU_{b,1}$, with the nuclear bath propagators $U_0$ and $U_1$ in the respective 0 and 1 electron manifolds.   This can be understood as the nuclear bath measuring an electron superposition $\ket{+}_e = \frac{1}{\sqrt{2}}\left(\ket{0} + \ket{1}\right)$;  the coherence of the electron is determined by the distinguishability of the nuclear evolutions in the different halves of the superposition. It is straightforward to show \cite{Maze:2008p1637} that for the spin-echo sequence with $\rho_{nuc.} = \openone$, the expectation value of the superposition state $\sigma_+  = \ketbra{+}{+}$ decays as $ \langle\sigma_+\rangle = \frac{1}{2} + \frac{1}{2} Re\left\{Tr\left((U_0U_1)^\dagger(U_1U_0)\right)\right\}$.  If the hyperfine quantization axes are the same, then $U_0$ and $U_1$ commute, and there is an echo at all times.  However, when the quantization axes are not co-linear, then there are additional echo modulations.  The initial echo decays on a timescale of a few $\mu s \approx 1/\sqrt{\sum_jA_j^2}$.  We can suppress this decay by rapidly switching between $U_0$ and $U_1$ (with electron $\pi$ pulses) so that they effectively commute - as in a Trotter expansion.  At longer times, there are unique circumstances where there is a full echo revival.  In the $\ketbra{0}{0}_e$ subspace, where there is no hyperfine interaction, $U_0$ is dominated by the Zeeman term; hence, it is the same for all $^{13}C$ nuclei and $U_0 = \exp(-it(\sum_j\omega_jI_z^j + \Ham_{dip.})  \approx \openone$ at $t = n\Omega_L$, integer multiples of the Larmor period, $\Omega_L$.  Since $U_0$ will then factor out of the trace expression, there are full echo revivals limited by $\Ham_{dip.}$ and any off-axis static magnetic field \cite{L.Childress10132006,Maze:2008p1637,Stanwix:2010p2867} (Figure \ref{RevivalsExpData} (a)).  We can suppress this decay too by adding more pulses; however,  the minimum pulse spacing must be a full Larmor period and the pulse spacings must be commensurate to achieve an echo. 
\begin{figure}[htbp]
\begin{tabular}{cc}
\includegraphics[scale=0.5]{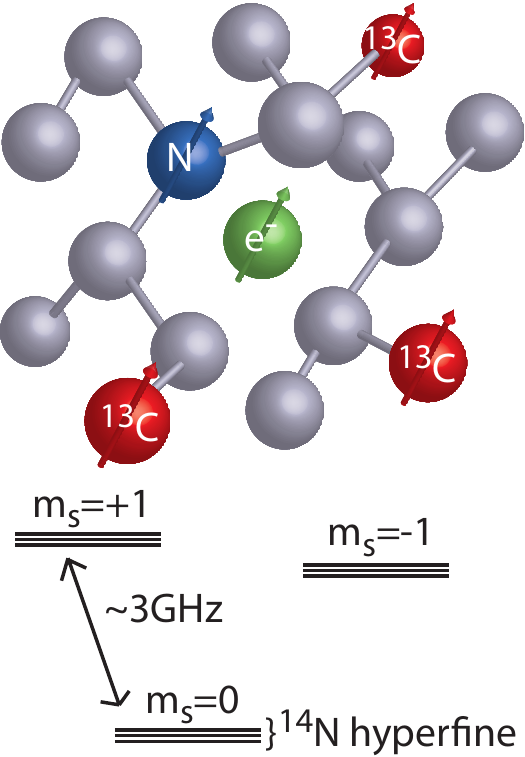} &
\includegraphics[scale=0.55]{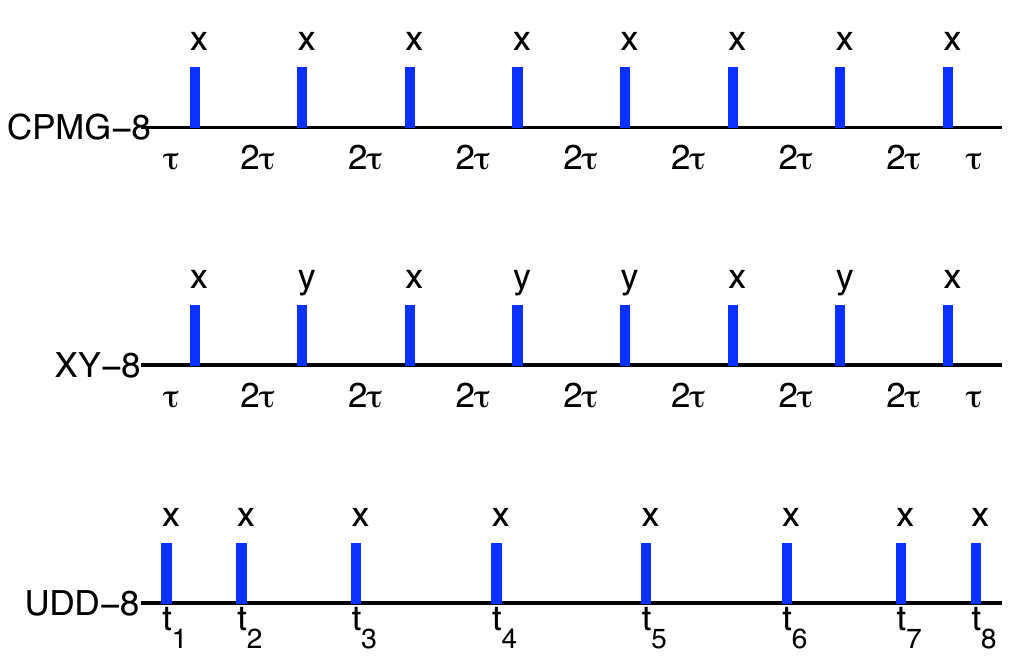} 
\end{tabular}
\caption{\label{CrystalandPulseSpacings}  NV in a diamond lattice and electron ground state level diagram.  The electron spin (green) is coupled to the surrounding nuclear spins: $^{14/15}N$ (blue) and $^{13}C$ (red) in the lattice.   An example with eight pulses of the three decoupling sequences used: CPMG; XY-8 and UDD.  CPMG and XY have the same pulse spacings but the phases of the XY-8 pulses are alternated.  The timing of the centre of the $k^{th}$ UDD pulse with $N$ total pulses is given by $\sin^2(\pi k/(2N+2))$.  }
\end{figure}

Multiple periodic control pulses have been used to suppress fluctuating fields since the beginnings of magnetic resonance \cite{Carr:1954p1927}.  Its experimental usefulness has been demonstrated in the context of QIP for nuclear bath hyperfine noise \cite{Bluhm:2010p2084,Du:2009p1299} and in ion traps \cite{Biercuk:2009p846}.  Of late, there has been interest in potential improvements offered by non-periodic spacing of the decoupling pulses.   Uhrig dynamical decoupling (UDD) \cite{Uhrig:2007p3053}, a variant of gradient moment nulling in NMR\cite{Keller:1988p3065}, is provably optimal \cite{Yang:2008p2386} in certain circumstances.   Specifically, UDD is optimal in a time-expansion of the unitary propagator when the noise has a sharp high-frequency cutoff.  When the cutoff is relatively softer, conventional CP spacing is preferred \cite{Pasini:2010p2307,Lee:2008p443}.   Intriguingly, while suppressing unwanted fluctuations, both approaches can be used to improve the sensitivity of NV magnetometers \cite{Hall:2010p1780,Taylor:2008p1323}.

The noise model for a central spin with varying anisotropic hyperfine couplings is not conventionally studied theoretically, although its pernicious effects on control fidelity have been noted \cite{Saikin:2003p3057}.  We show experimentally that for this model, equally spaced CP style sequences perform no worse than the optimized pulse spacings of UDD.   These experimental tests provide relevant results for the realistic case where there are losses due to pulse imperfections.   However, this loss of fidelity from pulse errors is dramatically state dependent.  For example, states along the rotation's axis are relatively insensitive.  However, for QIP applications the input state may be unknown;  this necessitates compensated sequences robust to pulse errors.  We investigated three decoupling sequences (Figure \ref{CrystalandPulseSpacings}): conventional CPMG, which for even numbers of pulses compensates for pulse errors along the direction of the rotation axis; the XY family which are compensated for all three axes of the Bloch sphere \cite{Gullion:1990p1244}; and finally the UDD which is uncompensated for pulse errors. 

We used a standard purpose-built confocal microscope setup for optical initialization and readout of single NV centres in single crystal diamond.  We controlled the effective two level system ($m_s=0,1$) with microwaves modulated by an arbitrary waveform generator to provide arbitrary microwave amplitude and phase \footnote{See Supplementary Material}.  

\begin{figure}[htbp]
\begin{tabular}{c}
\includegraphics[scale=0.55]{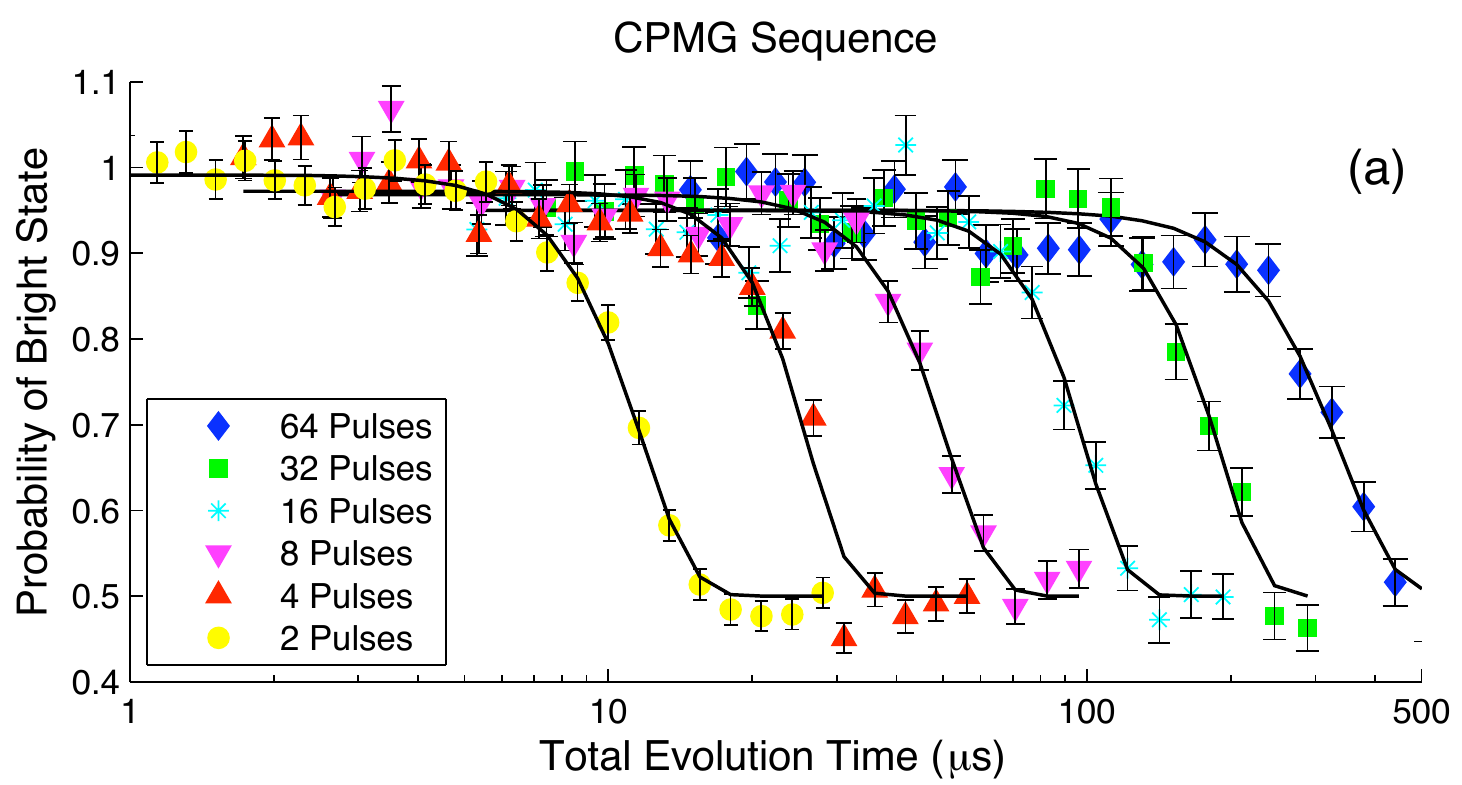} \\
\includegraphics[scale=0.55]{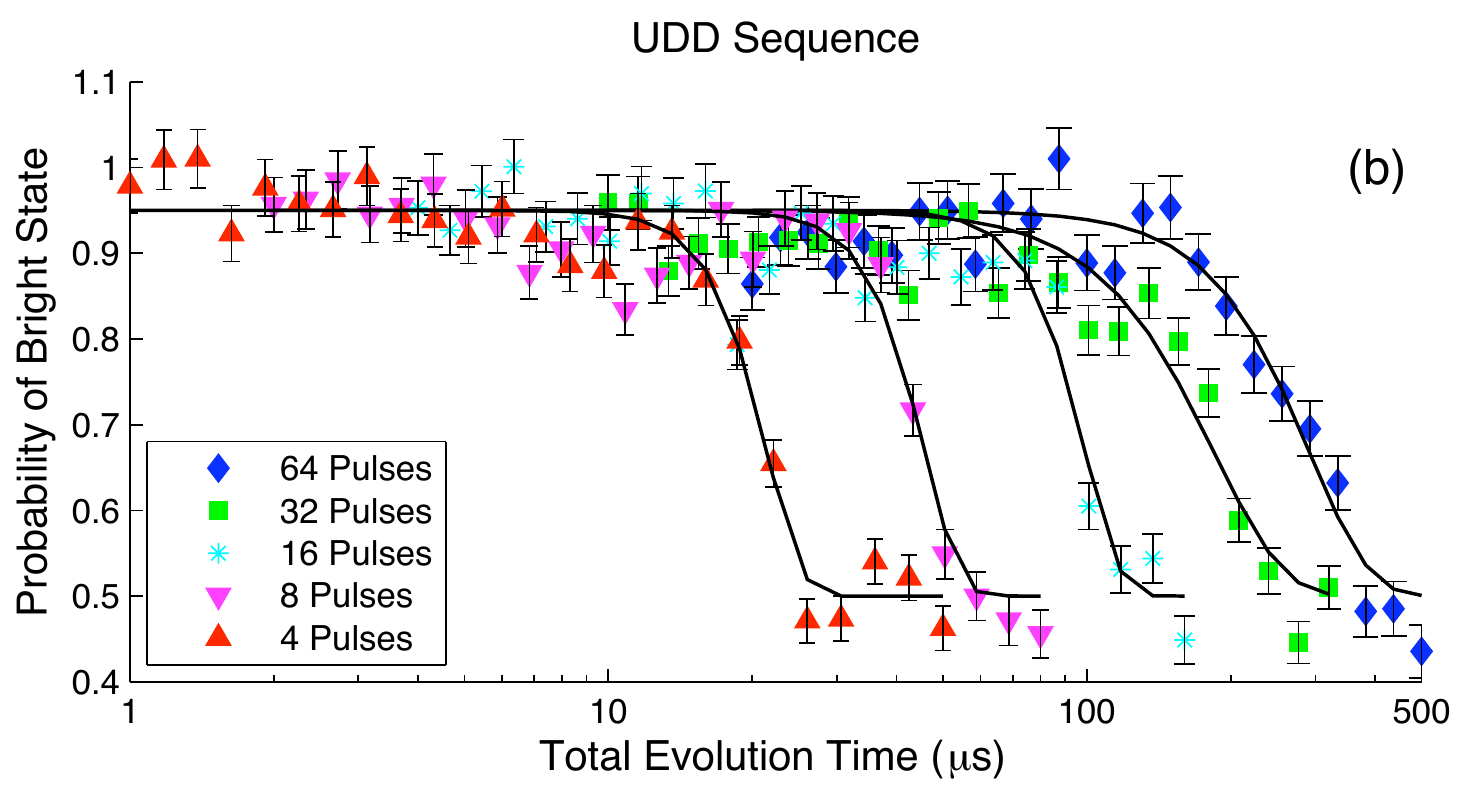} \\
\includegraphics[scale=0.55]{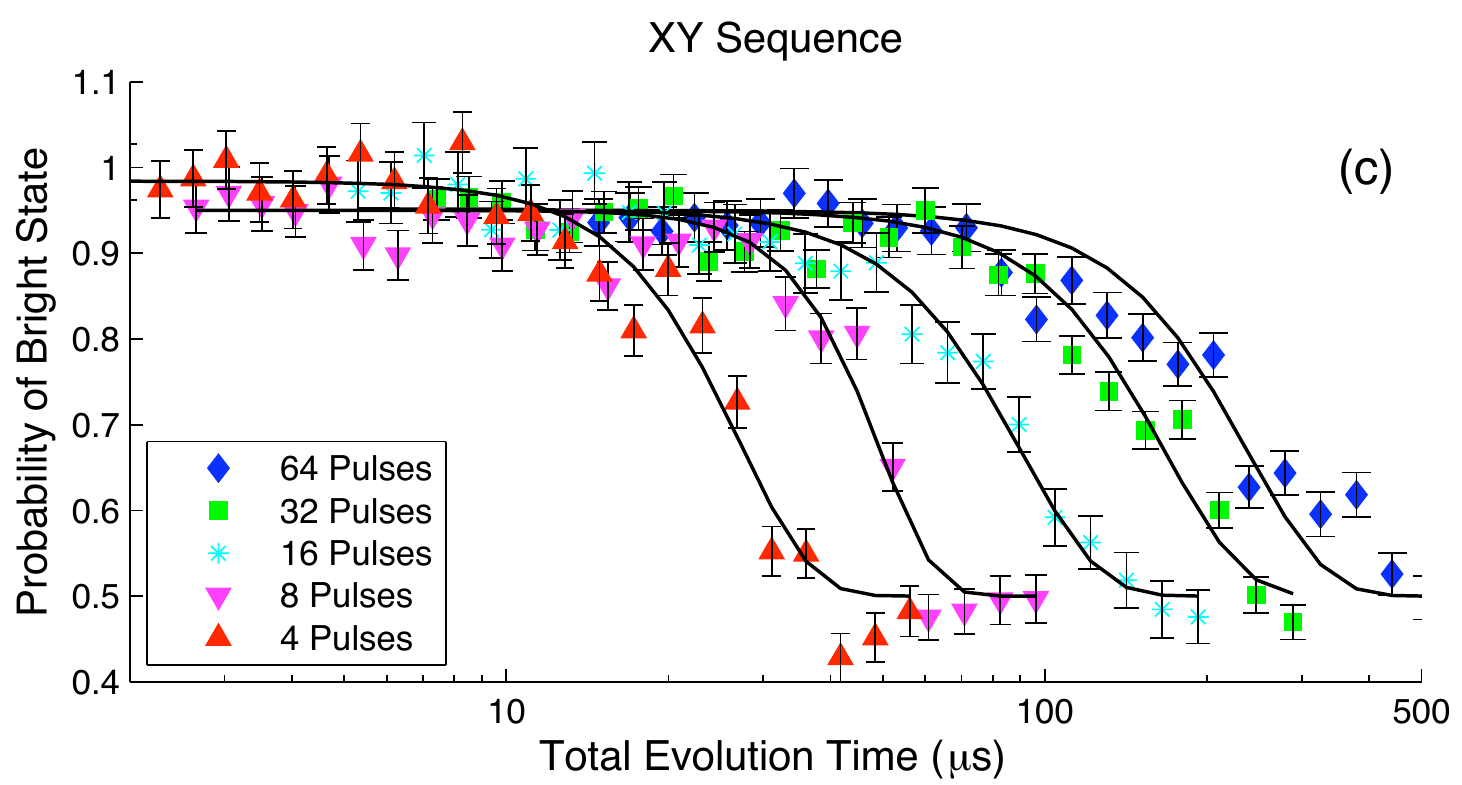} 
\end{tabular}
\caption{\label{ExpData} Short-time coherence decay for states along the pulse rotation axis with (a) CPMG, (b) UDD and (c) XY sequences.   Fits are to the phenomenological  form $s(t) = 0.5 + A\exp(-(\frac{t}{T_2})^k)$ where $3 < k < 6$.  Error bars are propagated from photon counting statistics. }
\end{figure}

When the input coherence is parallel to the pulse rotation axis  we can explore the extension in $T_2$ with the number of pulses and the effect of variable pulse spacing.  However,  the UDD and XY sequences still suffer from pulse sequence imperfections and there is evidence of an additional early decay with more than 16 pulses.  The experimental results are shown for the initial short-time coherence decay in Figure \ref{ExpData} and the extracted effective $T_2$'s in Table \ref{T2TableExp}.  

\begin{table}[htbp]
\caption{\label{T2TableExp}  Effective $T_2$ ($\mu s$) extracted from the fits to the curves in Figure \ref{ExpData}.   For 2 pulses CPMG and UDD are equivalent and there is no identity XY sequence possible.  Experiments with fixed pulse spacing and going out to 1000's of pulses indicate   }
\begin{ruledtabular}
\begin{tabular}{cccc}
\# of Pulses & CPMG & UDD & XY  \\
2	&	$11.8\pm0.4$	&	$11.8\pm0.4$	&	-	\\ 
4	&	$26\pm2$		&	$22\pm2$		&	$27\pm2$	\\ 
8	&	$51\pm3$		&	$46\pm3$		&	$50\pm3$	\\ 
16	&	$100\pm5$	&	$99\pm8$	&	$91\pm6$	\\ 
32	&	$190\pm13$	&	$185\pm17$	&	$168\pm12$	\\ 
64	&	$340\pm25$	&	$293\pm25$	&	$239\pm21$	\\ 
\end{tabular}
\end{ruledtabular}
\end{table}

We posit that the observed performance of the sequences is due to the nature of the non-Markovian spin bath inherent to NV centres in diamond. For the NV defect at low fields, the aniostropic hyperfine, and associated coherent dynamics from the dilute spin bath is an important non-Markovian noise source not easily treatable as a semi-classical fluctuating magnetic field at the electron spin \cite{Sousa:2009p1278}.  We validate our assumption via simulations of the NV centre and nuclear spin bath, made tractable through the disjoint cluster method  \cite{Maze:2008p1637}.  These simulations provide insight and confirmation of the validity of our experimental results \footnote{See Supplementary Material}.  They also confirmed the surprising almost linear increase in $T_2$ as opposed to the expected $n^{\frac{2}{3}}$ dependence for a semi-classical field \cite{Sousa:2009p1278}.  They also confirmed that the UDD underperformance is not caused by imprecision in the pulse spacings due to the finite AWG clock frequency, where experimentally the timings were rounded to the nearest 2 ns \cite{Biercuk:2009p2305}.   

\begin{figure}[htbp]
\includegraphics[scale = 0.55]{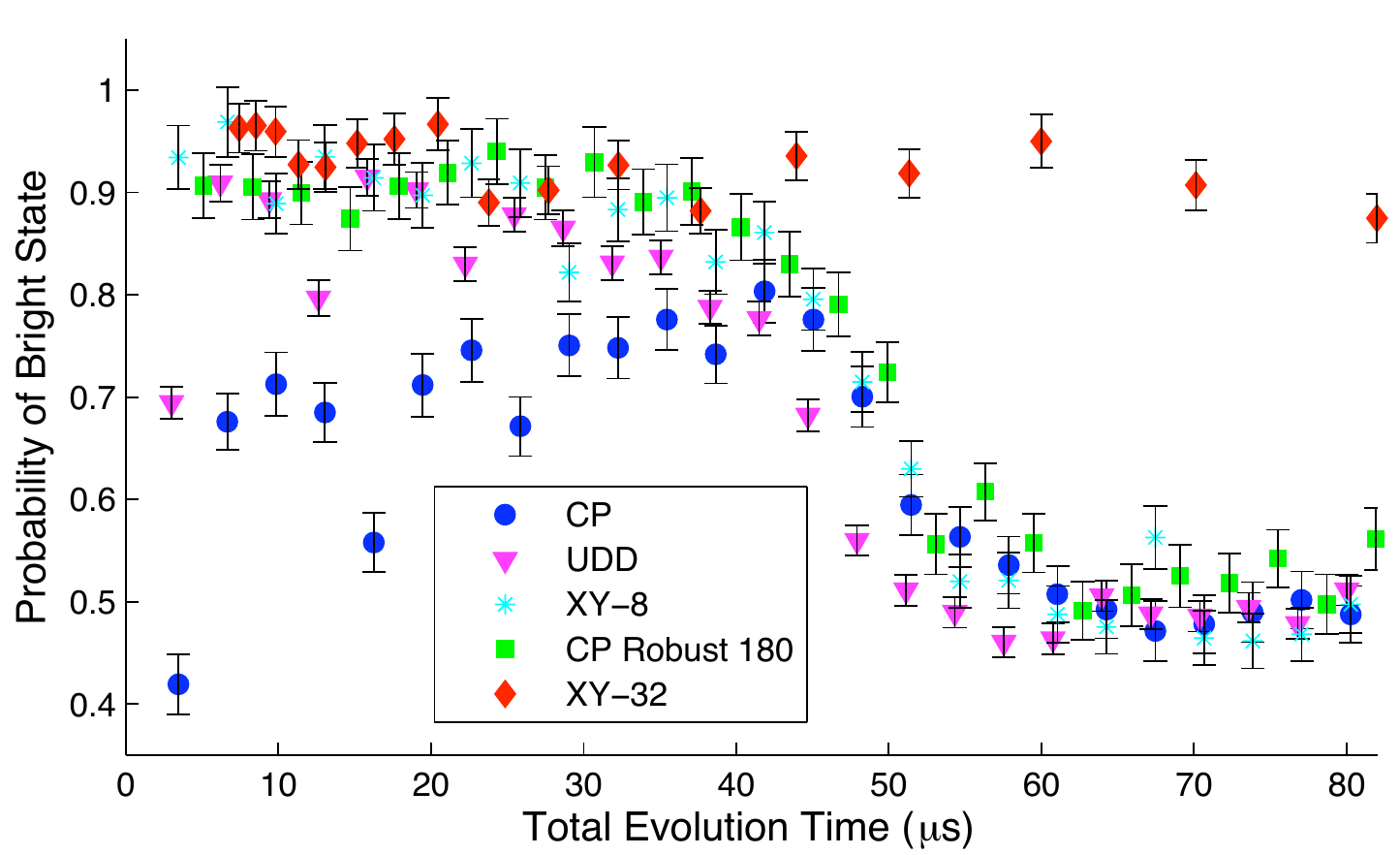}
\caption{\label{YInputStateFig} CP-8 (circles), and UDD-8 ($\blacktriangledown$)  with the input state perpendicular to the pulse rotation axis.   The deleterious effect of the coherent pulse errors leads to two time scales for the decay: initial coherent oscillations are rapidly damped as they are not properly refocussed and then same long time decay as the parallel state.    Pink ($\blacktriangledown$) Simply modulating the phases of the pulses as in the XY-8  sequence  ($\bigstar$) can overcome this.  Robust composite pulses  ($\blacksquare$) provide similar relief despite having five times the number of pulses.  Also shown is the XY-32 sequence ($\diamond$) demonstrating that we can preserve an arbitrary state to much longer times by shortening the pulse spacing.  }
\end{figure}

An additional complexity in the NV system is the presence of the $^{14}N$ hyperfine coupling of ~2.1MHz.  This leads to substantial coherent off-resonance errors in the $\pi$ pulses.  Our single 32 ns, nominally square, pulses have a simulated fidelity of 98.6\% (assuming a maximally mixed $^{14}N$ state).  Any experimental implementation will fall short of this due to errors such as fluctuations in the microwave power or transient variations in the phase of the pulse at the rising and falling edges.  For multiple pulse sequences where all the pulses are about the same axis, these errors add coherently and rapidly destroy the effectiveness for particular input states.   For our setup, beyond 4 pulses, decoupling for states perpendicular to the rotation axis was ineffective (Figure \ref{YInputStateFig}).    However, for QIP applications, the input state is unknown and the decoupling sequence should be robust to all input states.  

One approach is to make the decoupling sequence itself more robust  for any initial state: the motivation for the XY family of sequences \cite{Gullion:1990p1244}.  By alternating the phase of 180 degree pulses about the $\pm X$ and $\pm Y$ axes and concatenating appropriately the pulse sequence can be compensated for pulse errors and the overall errors made \emph{isotropic} (Figure \ref{YInputStateFig}).   Indeed, our experimental results show that this sequence provides useful refocussing for both input states where sequences with all the pulses about the same axis are not useful.  The tradeoff is that this sequence will not perform as well as CPMG for a known input state (Figure \ref{ExpData}). 

A second approach is to use composite pulse sequences to make the net rotation of individual pulses more robust \cite{Borneman:2010p3048}.  We chose the $180_{30}-180_{0}-180_{90}-180_{0}-180_{30} = Z_{60}180_{0}$ sequence for its robustness to resonance offsets and ease of calibration \footnote{This pulse is attributed to Dr E. Knill.}.  The additional $Z$ rotation is easy to absorb in an abstract reference frame shift.  Implementing this composite sequence with 46ns Gaussian pulses (to avoid overlap of transients at the pulse edges) gives a ideal fidelity of 99.95\%, and as seen in Figure \ref{YInputStateFig}, an overall improved performance.  Combining composite pulses with XY sequences provides even better performance. 

The effect of the multiple pulse echoes can be particularly dramatic when we observe the echo revivals.  These revivals will be useful in refocussing the electron spin coherence while performing nuclear gates in QIP applications \cite{Cappellaro:2009p1554}.   The revivals decay due to off-axis fields and nuclear dipole-dipole coupling of up to a few kHz.  By applying more than one echo pulse, then we can suppress the decay of the revivals at the expense of less frequent revivals: the revivals only occur when the shortest pulse spacing corresponds to a nuclear identity operation in the $m_s = 0$ manifold.  Hence, it is also not possible to see echo revivals with the unequal pulse spacing of UDD.   We were able to demonstrate an over seven-fold increase in the ``$T_2$" over the single spin echo time by using the robust XY-4,XY-8 and XY-16 sequences  from $220 \mu$s to over 1.6 ms (Figure \ref{RevivalsExpData}).  This newly revealed extraordinary coherence time is then comparable to the 1.8 ms reported for isotopically purified diamond, some of the longest room-temperature coherence times for a solid state electron spin \cite{Balasubramanian:2009p1632}.

In summary, we have experimentally demonstrated dramatic increases in the effective dephasing time of a nitrogen vacancy centre in diamond by using robust sequences and composite pulses to suppress both the fluctuations that lead to the dephasing and the intrinsic errors in the pulses themselves.  We expect these robust sequences will prove useful not only in both NV magnetometry and QIP applications but also in other solid-state QIP implementations such as quantum dots \cite{Bluhm:2010p2084}  or superconducting qubits \cite{Cywinski:2008p2864}.  The single spin echo coherence time of our diamond was not particularly long at 220 $\mu s$, as coherence times of up to $600 \mu s$/1.8 ms in $1.1\%/0.3\% ^{13}C$ have been reported.  These sequences should similarly enhance the coherence times of these approaches and help them approach the ultimate $T_1$ limit.        
 
We thank P. Cappellaro for encouraging discussions.  This work was supported in part by NSF, DARPA (QuEST), the National Security Agency under Army Research Office (ARO) contract number W911NF-05-1-0469, and in part by the Canada Excellence Research Chairs program and NSERC.    Both the Deflt and Stuggart groups have very recently released complementary similar results \cite{Lange:2010p3071,Naydenov:2010p3068}.

\bibliography{Colm_Papers.bib}

\section{Supplementary Materials}

\subsection{Experimental Methods}
We used a purpose-built confocal microscope setup for imaging single NV centres in a commercial (Element 6, electronic grade, CVD-grown, $< 5$ppb native N impurities) single crystal diamond.  The density of centres and photon counting rates strongly indicate we were focussed on a single NV, athough we did not explicitly confirm this through photon correlation measurements.   Microwaves from a low phase-noise source (Agilent E8241A) are brought to the centre via a $25 \mu$m diameter copper wire mounted on top of the crystal.  The microwaves were single-side-band modulated at an IQ mixer (Marki Microwave IQ0255LMP) by a 100MHz I.F. created by a 500MHz AWG (Tektronix AWG5002) to allow arbitrary amplitude and phase control.  This I.F. approach allows us to leverage the 14-bit vertical precision of the AWG to set both the phase and amplitude of the microwaves.  Non-linearites in the amplitude were corrected for by measuring the spin-dynamics at a range of power levels and fitting a non-linear correction curve:  $\langle\sigma_z\rangle$ was measured as the power was increased from 1 to 100\% and the Rabi oscillation curve was heuristically fit to create a non-linear correction table. An 10W microwave amplifier (HD Communications HD20323  ) gave nutation frequencies of 15MHz.     The $T_2^*$ of the centre measured from fitting a Gaussian decay to Ramsey fringes was $2.7 \mu$s; the $T_2$ measured in a single spin echo was $220 \mu$s and the $T_1$ for the $m_s=0$ electron state was  6 ms.    The measured fluoresence (averaging 0.015 photons/350 ns counting period for the $m_s = 0$ sublevel) was referenced and scaled.  Before each individual run of the experiment, references were taken for both the bright state (a 3 $\mu$s delay after optical pumping) and the dark state (using a robust broadband adiabatic state-to-state inversion pulse to the $m_s = +1$ state).    A static magnetic field of a few tens of Gauss was applied along the NV axis using a 3-axis Helmholtz coil setup.  The field was aligned by maximizing the Zeeman splitting as measured through CW ESR spectra and comparison of the revival frequency in the spin-echo experiment with with the Zeeman shift of the electron resonance frequency.   This static field splits the $m_s=\pm1$ levels;  we concentrated on the $m_s = 0 \rightarrow +1$ transition at 2.98 GHz. 

\subsection{Disjoint-Cluster Simulations}

As shown in the main text, the electron spin dephasing is determined by the distinguishability of the nuclear spins' evolution.   Simulating this enormous Hilbert space of 100's of nuclear spins can be made tractable by breaking the system up into clusters of strongly coupled nuclear spins: the so-called disjoint-cluster method.  The code was written in Matlab and is available from the authors.  10 lattice sites in every direction from the defect were considered and $^{13}C$ spins were randomly distributed at 1.1\% giving approximately 800 $^{13}C$ spins.    The nuclear clusters were considered with up to 6 $^{13}C$ spins and the full second order perturbation Hamiltonian accounting for electron mediated nuclear-nuclear interactions and the hyperfine-enhanced nuclear Zeeman effect was calculated.   The simulations only consider perfect zero-time pulses and so we can only explore the effect of pulse spacing: CP equal time spacing versus UDD optimized spacing.   

\begin{figure}[htbp]
\begin{tabular}{c}
\includegraphics[scale=0.55]{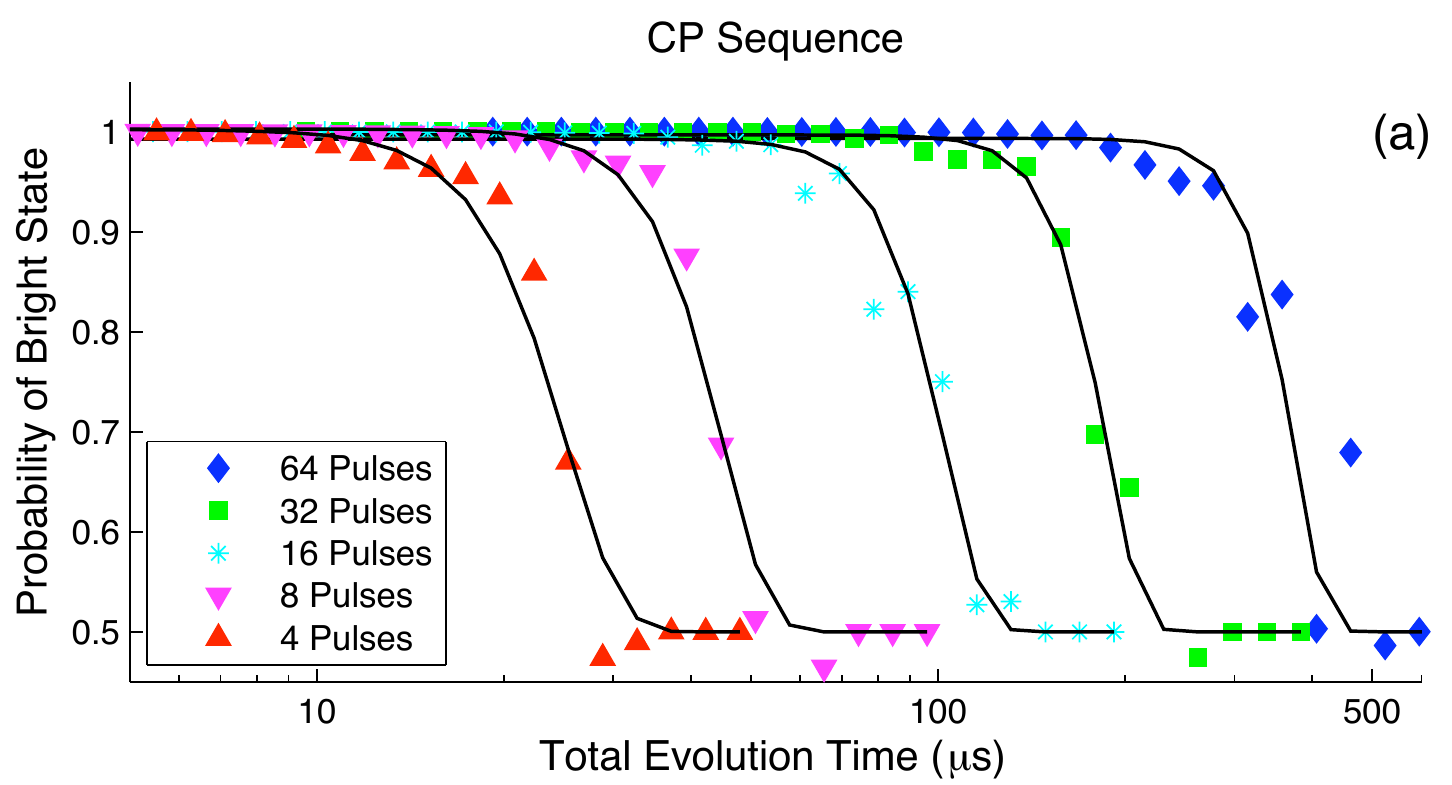} \\
\includegraphics[scale=0.55]{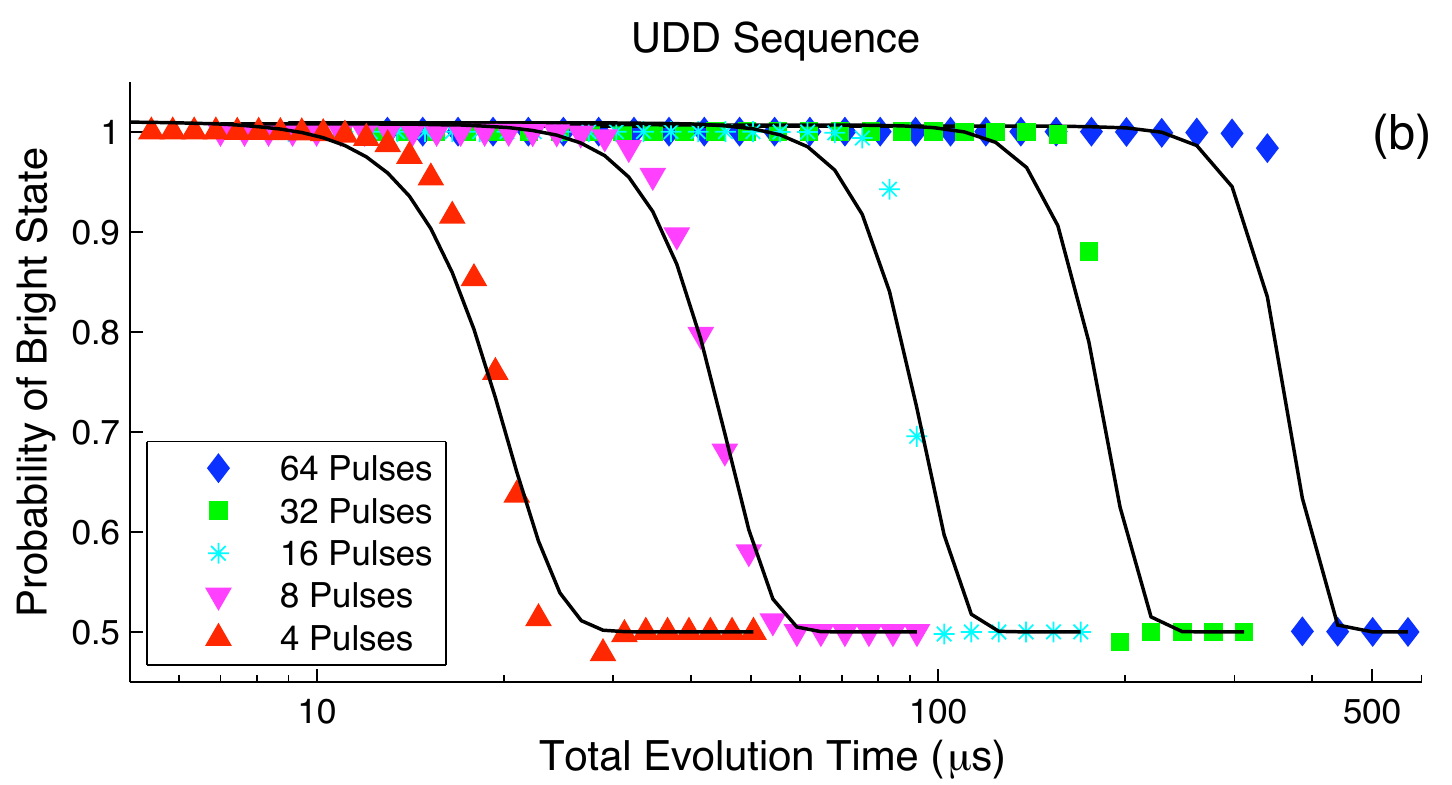} 
\end{tabular}
\caption{\label{NumericalShortTimeData} Sample decay curves for a particular $^{13}C$ distribution from the disjoint cluster method for (a) CP style pulse spacings and (b) UDD pulse spacings.    }
\end{figure}

Not surprisingly, the specific form of the decay curves varied significantly from run to run, depending on the different random distributions of $^{13}C$ spins around the simulated defect.  The sample curves shown in Figure \ref{NumericalShortTimeData} all correspond to a simulation with a particular defect with no stronghyperfine interactions  ($> 300$ kHz).  This is similar to the experimental situation where certain centres are avoided because of their unresolved hyperfine interactions lead to a rapid decay and broad CW lines.   The simulations confirmed the experimental CP-style equal pulse spacing out-performs or equals the performance of UDD, the scaling of $T_2$ with the number of pulses and also showed remarkable agreement with the experimental results.

\begin{table}[htbp]
\caption{\label{T2TableSim}  Effective $T_2$ ($\mu s$) extracted from the fits to the simulated curves in Figure \ref{NumericalShortTimeData}.      }
\begin{ruledtabular}
\begin{tabular}{ccc}
\# of Pulses & CPMG & UDD  \\
2	&	$11.1\pm0.3$	&	$11.1\pm0.3$	\\ 
4	&	$25\pm1$		&	$20\pm1$		\\ 
8	&	$45\pm1$		&	$46\pm7$		\\ 
16	&	$103\pm3$	&	$95\pm3$	\\ 
32	&	$187\pm7$	&	$189\pm7$	\\ 
64	&	$374\pm16$	&	$374\pm14$	\\ 
\end{tabular}
\end{ruledtabular}
\end{table}

\begin{figure}[htbp]
\includegraphics[scale=0.55]{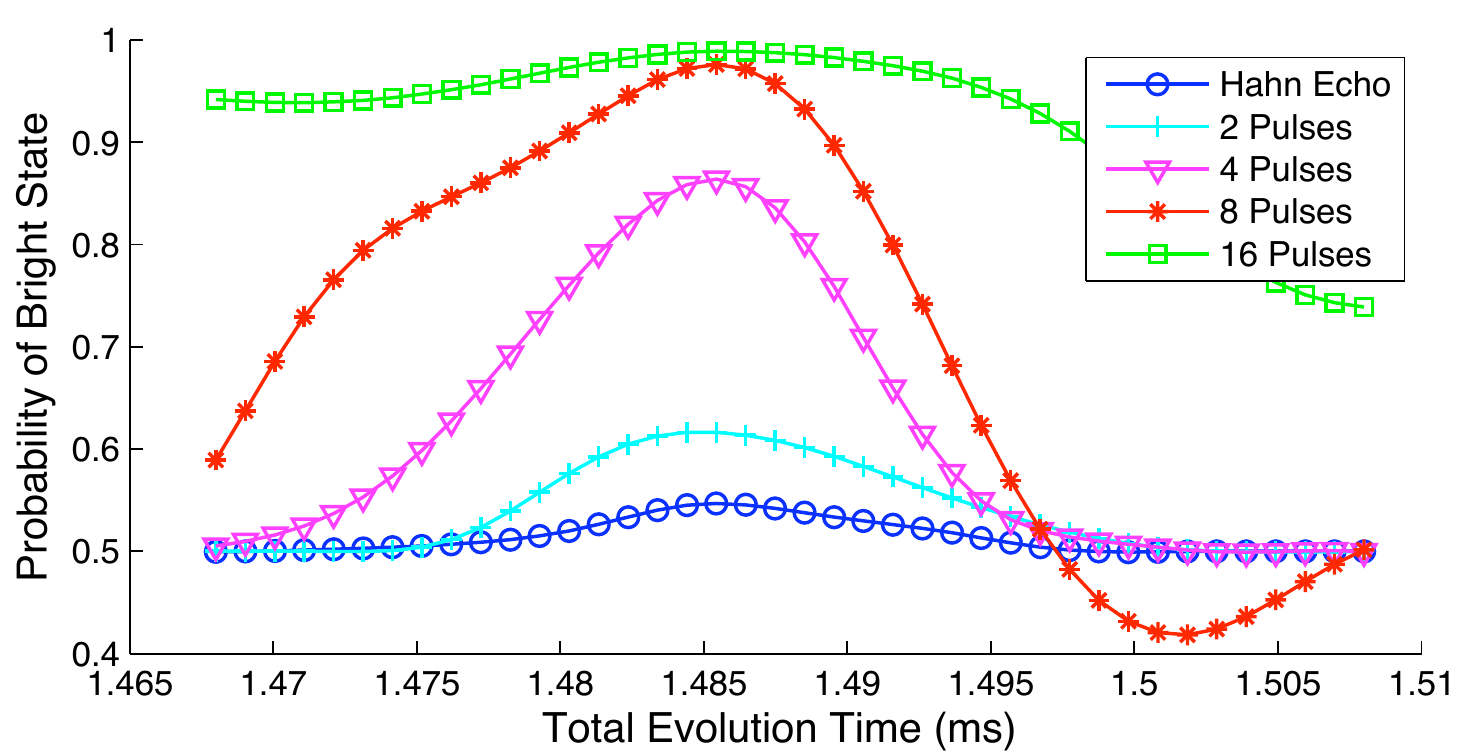} 
\caption{\label{NumericalLongTimeData} Sample revival curves for multiple pulse echos showing the expected improvement with additional pulses.  Whereas the single pulse echo has decayed to almost nothing, 16 pulses can create an almost full revival.  }
\end{figure}

We can also use the simulations to look at the long-time echo revivals and how they are affected by multiple pulses.  Figure \ref{NumericalLongTimeData} shows the revival at 32 Larmor periods, the longest we looked at experimentally.  It shows both the increase in fidelity with increasing numbers of pulses as expected, but tends to overestimate the observed coherence times.  Perhaps this is due an additional decay mechanism though other nearby defects or a slightly mis-aligned magnetic field.  

\section{Fixed Pulse Spacing CPMG}
The conventional approach to CPMG from NMR is to fix the pulse spacing and increase the number of pulses rather than fixing the number of pulses and increasing the time.  This allows us to explore the limit of our decoupling approach for certain input states.  The results show that at short pulse spacings we get coherent oscillations in the decay as we rapidly modulate the dynamics but we extend the coherence time out to greater than 800 $\mu$s.  At longer pulse spacing we get a more conventional exponential decay as there is sufficient time for small but irreversible dynamics in the nuclear bath.    At even longer pulse spacings there is no high-fidelity echo for even one pulse due to the rapid ESEEM decay.

\begin{figure}[htbp]
\includegraphics[scale=0.8]{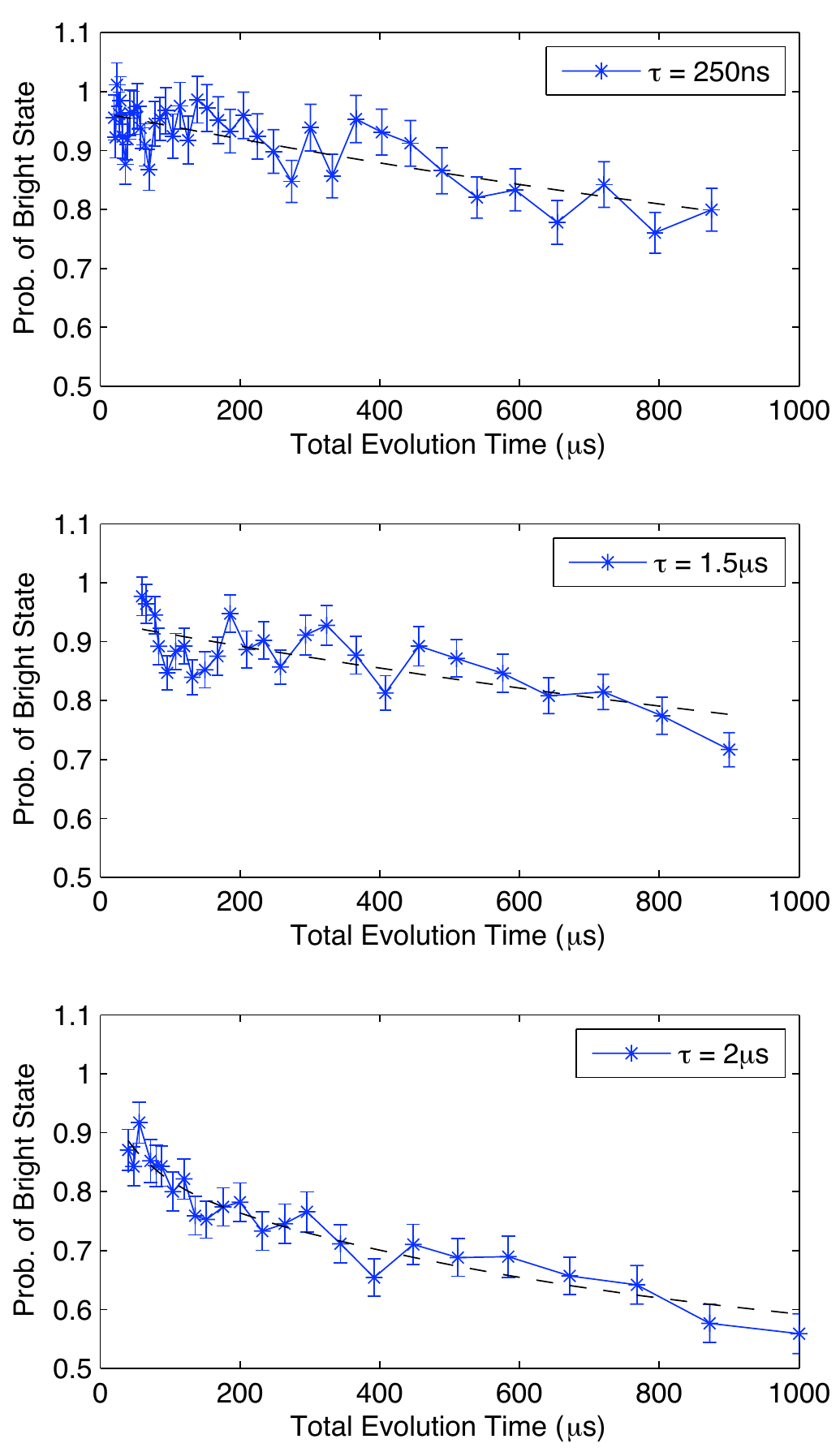} 
\caption{\label{CPMGFixedSpacing}  CPMG decay with fixed pulse spacings.  $\tau$ refers to the initial delay in the $(\tau-\pi-2\tau-\pi-\tau)^n$ sequence. Fits are to an exponential decay but the quality of the fits is not high enough to give a decay constant.    }
\end{figure}

\end{document}